\documentclass{iopart}
\usepackage{amsfonts} 
\usepackage{latexsym} 
\usepackage{amssymb} 
\usepackage[american]{babel}

\def\II{{\mathbb I}} 
\def\RR{{\mathbb R}}

\def\ZZ{{\mathbb Z}}

\def\tr{\mathrm{ tr\,}}

\def\Det{\mathrm{ Det\,}}

\def\vol{\mathrm{ vol\,}}

\def\diag{\mathrm{diag\,}} 
\def\Spin{\mathrm{Spin}}

\def\be{\begin{equation}} 
\def\ee{\end{equation}} 
\def\bea{\begin{eqnarray}} 
\def\eea{\end{eqnarray}} 
\def\bed{\begin{definition}{\ }}
\def\eed{\end{definition}}
\def\bd{\begin{description}}
\def\ed{\end{description}}
\def\bc{\begin{center}}
\def\ec{\end{center}}

\begin{document}

\title[Thermal Yang-Mills Theory In the Einstein Universe]
{Thermal Yang-Mills Theory In the Einstein Universe}

\author{ Ivan G Avramidi and Samuel Collopy}

\address{
 Department of Mathematics,
New Mexico Institute of Mining and Technology,
Socorro, NM 87801, USA}
 \eads{\mailto{iavramid@nmt.edu}, \mailto{samuel.collopy@gmail.com}} 

\begin{abstract}

We study the stability of a non-Abelian chromomagnetic vacuum in Yang-Mills
theory in Euclidean Einstein universe $S^1\times S^3$. 
We assume that the gauge group is a simple compact group $G$ containing
the group $SU(2)$ 
as a subgroup and consider static covariantly constant gauge fields on $S^3$ taking values
in the adjoint representation of the group $G$ and forming a representation of the
group $SU(2)$. We compute the heat kernel for the Laplacian acting on fields 
on $S^3$ in an arbitrary representation of $SU(2)$ and use this result to compute
the heat kernels for the gluon and the ghost operators and the 
one-loop effective action.
We show that the only configuration of the covariantly constant
Yang-Mills background that is stable is the one that contains only spinor
(fundamental) representations of the group $SU(2)$; all other configurations contain
negative modes and are unstable. For the stable configuration we
compute the asymptotics of the effective
action, the energy density, the entropy and the heat capacity
in the limits of low/high temperature and small/large volume and show that the energy density
has a non-trivial minimum at a finite value of the radius of the sphere $S^3$.

\end{abstract}



\section{Introduction}
\setcounter{equation}0

A deep understanding of the physics of quantum Yang-Mills gauge field theory at low
energies is lacking due to the failure of perturbation theory. At low
energies the interaction becomes strong which leads to the
phenomenologically observed property of confinement in quantum chromodynamics.
However, the precise nature of the physical mechanism responsible for
confinement is still not known. From the field-theoretical point of view this
means that the vacuum of Yang-Mills theory at low energies has a far more
complicated structure than the perturbative one. One of the models of a
non-perturbative Yang-Mills vacuum is the chromomagnetic vacuum that has been
proposed by Savvidy \cite{savvidy77}. Savvidy considered a constant $SU(2)$
gauge field (which was necessarily Abelian) in $4$-dimensional flat spacetime
and showed that the trivial zero-field perturbative vacuum is unstable under
creation of a constant chromomagnetic field. Further, Nielsen and Olesen
\cite{nielsen78} showed that the Savvidy vacuum is itself unstable, meaning
that the physical nonperturbative vacuum has an even more complicated structure.
It has been suggested \cite{nielsen79} that the real vacuum is likely to have a
small domain structure with random constant chromomagnetic fields.

One way to stabilize the chromomagnetic vacuum is to increase the dimension of
the space-time. In our papers \cite{avramidi95a,avramidi99} we studied the
chromomagnetic vacuum (for general compact gauge groups) in higher dimensions in
flat spacetime and showed the existence of stable chromomagnetic configurations
in dimensions greater than four. Another way to stabilize the vacuum is to
consider curved manifolds. It is easy to see that a positive space curvature can
provide an effective mass term for the gauge fields on the chromomagnetic
vacuum, thus, making the vacuum stable. Of course, on curved manifolds the
notion of constant chromomagnetic gauge fields has to be replaced with {\it
covariantly constant fields}. 

In our recent paper \cite{avramidi12} we
considered the finite temperature Yang-Mills theory on $S^1\times S^1\times S^2$
with an {\it Abelian covariantly constant background} on $S^2$. We showed that
despite the positive spatial curvature the gluon operator still has negative
modes for any compact semi-simple gauge group, which means that the vacuum with 
covariantly constant chromomagnetic fields on $S^2$ does not represent the true
vacuum of Yang-Mills theory.
This happens because 
{\it on spheres, and, more generally, on symmetric spaces, 
covariantly
constant gauge fields cannot have an arbitrary value of the chromomagnetic
field independent on the spatial curvature}; 
there are severe algebraic constraints that force the {\it strength of the
chromomagnetic
field to be of the same order as the spacial curvature}. This means that the {\it stabilizing
role of the spatial curvature is depressed by the destabilizing role of the
chromomagnetic field}, even on curved manifolds with constant positive curvature.

The present paper is a continuation
of the study of Yang-Mills theory on spheres; our primary  
goal here is to extend this analysis to the Einstein universe
$S^1\times S^3$ with a {\it non-Abelian covariantly constant background} on $S^3$
with a compact simple gauge group $G$ that has the group $SU(2)$ as a subgroup.
We refer to the paper \cite{avramidi12} for general introduction
and notation.

This paper is organized as follows. In Sect. 2 we briefly describe our model and fix notation.
In Sect. 3 we derive the one-loop effective action in terms of the heat kernels of relevant operators.
In Sect. 4 we compute the heat kernel on $S^1$. In Sect. 5 we describe in detail the geometry of $S^3$
and compute the scalar heat kernel on the group $SU(2)$, which is used later for the calculation of the 
heat kernel for arbitrary fields on $S^3$ in Sect. 6. 
In Sect. 7 we use these results to compute the 
trace of the Yang-Mills heat kernel (which is the difference between the gluon heat kernel and the ghost one)
and the effective action.

The {\it main results of this paper} are the calculation of the heat kernel for arbitrary fields
on $S^3$ and the proof that 
the {\it minimal eigenvalue of the gluon operator is positive only in a very specific case
when the representation of the gauge fields does not contain any higher spin ,
$j\ge 1$, representations
of the group $SU(2)$ but contains only  the spinor (fundamental) representation, $j=1/2$, of $SU(2)$}.
In all other cases, that is, {\it when the representation of the gauge fields
contains at least one representation of $SU(2)$ with $j\ge 1$, 
the minimal eigenvalue is negative and the 
heat kernel grows exponentially at infinity 
leading to the infrared instability of the chromomagnetic vacuum}. 
Of course, the gauge fields are taken in the adjoint representation of the gauge group $G$.
The question whether the adjoint representation of a compact simple group $G$, that has
the group $SU(2)$ as a subgroup, can contain the spinor representation of $SU(2)$, is a 
representation-theoretic problem that we do not solve in this paper. 
If this is impossible then {\it our results indicate
the instability of any chromomagnetic background in Yang-Mills theory in Einstein universe}.

We assume that this is possible, that is,
the adjoint representation of the gauge group $G$ may contain 
the spinor representation of $SU(2)$, 
 and compute the heat kernel and the effective action for this specific case.
Of course, it does not make any sense to study thermodynamics of an unstable theory.
Therefore, in Sect. 8 we study thermodynamics of the model for this specific case; we
compute the entropy, the heat capacity and the pressure.

The quantum field theory on homogeneous spaces 
has an extensive bibliography. We list only some papers that had an influence on our own approach.
First of all, we would like to mention the early important papers by Dowker
\cite{dowker70,dowker71,  dowker72b, dowker72, dowker83} as well his papers with coworkers
\cite{dowker77, altaie78}.
A very good source is the excellent review by Camporesi \cite{camporesi90} and the 
reference therein  as well as the
papers \cite{rubin85,anderson90, camporesi92,camporesi94,camporesi96,elizalde96}.
A thermal Yang-Mills theory in Einstein universe was also studied in \cite{volkov96}.
More recent papers \cite{giombi08,david10,gopakumar11} have some overlap with our
work  since they also studied the heat kernel on $S^3$. 

We would like to stress from the very beginning
the differences with these papers. Our approach is completely different from the papers mentioned above;
it is based on a fundamental integral representation of the heat semi-group on homogeneous bundles
as an integral over the isotropy group
(see our previous work \cite{avramidi93,avramidi08,avramidi09}, for application of this approach to
quantum gravity see \cite{avramidi10a}). Also, while most of the authors deal with the heat kernel
for irreducible  (that is, traceless divergence free) tensor fields, we compute it for arbitrary fields. 

\section{Yang-Mills Theory}
\setcounter{equation}0

We consider the manifold $M=S^1\times S^3$ equipped with the standard product
metric $g_{\mu\nu}$ and
a compact simple gauge group $G$.
We denote the tensor components with respect to the coordinate basis
by Greek indices. 
Let $x^\mu$ be some local coordinates, ${\cal A}={\cal A}_\mu dx^\mu$ 
be the Yang-Mills connection one-form
taking values in the adjoint representation of the Lie algebra of the group $G$
and 
${\cal F}=\frac{1}{2}{\cal F}_{\mu\nu}\,dx^\mu\wedge dx^\nu
=d{\cal A}+{\cal A}\wedge{\cal A}$ be its curvature 2-form.
Then the  
classical action of the pure Yang-Mills theory is
\be
S=
\int_M d\vol\;
\frac{1}{2e^2}|{\cal F}|^2,
\ee
where $d\vol=dx\, g^{1/2}$  is the Riemannian volume element with $g=\det g_{\mu\nu}$,
$e$ is the Yang-Mills coupling constant and
$|{\cal F}|^2
=-\frac{1}{4}\tr g^{\mu\alpha}g^{\nu\beta}{\cal F}_{\mu\nu}{\cal F}_{\alpha\beta}$.

We choose a local orthonormal frame on the tangent bundle $TM$
and denote the tensor components with respect to the orthonormal frame by low case Latin indices
from the beginning of the alphabet.
The components of tensors on $S^3$ are denoted by Latin indices from the middle
of the alphabet.
We introduce the projection tensor $h_{ab}$ to the tangent space of $S^3$
and  the Levi-Civita symbol $\varepsilon_{abc}$ on $S^3$.

We assume that the background Yang-Mills curvature is parallel
$
\nabla_a{\cal F}_{bc}=0.
$
Such background on $S^1\times S^3$ has to satisfy the constraint
\be
[{\cal F}_{ab},{\cal F}_{cd}]=-\frac{4}{a^2}h^{[a}{}_{[c}{\cal F}^{b]}{}_{d]},
\ee
where $a$ is the radius of $S^3$.
To satisfy this constraint we assume that the gauge group $G$ 
has the group $SU(2)$
as a subgroup. 
Then there exist generators $X_i$ taking values in the adjoint representation of the
Lie algebra of the group $G$
satisfying the algebra $su(2)$
\be
[X_i,X_j]=-\varepsilon^k{}_{ij}X_k,
\ee
that is, $X_i$ form a representation $X$ of the algebra $su(2)$.
Now, the above constraint is satisfied by 
\be
{\cal F}_{ab}=\frac{1}{a^2}\varepsilon^i{}_{ab}X_i.
\ee
This means that the magnitude of a covariantly constant gauge field cannot be assigned
arbitrarily but it is determined by the radius of the sphere $S^3$, which also means that the
{\it spatial curvature and the chromomagnetic field are of the same order}, $1/a^2$.

In general, this representation is reducible. 
It decomposes into a sum of irreducible representations 
$
X_i=\diag\{Y^{j_1}_i\oplus\cdots\oplus Y^{j_N}_i\},
$
where $j_i$ are the standard non-negative 
integer or half-integer labels of the representations of $su(2)$.
The Casimir operator $X^2=X_i X^i$ is a diagonal matrix
\be
X^2
=\diag\left\{-j_1(j_1+1)I_{2j_1+1}\oplus\cdots\oplus -j_N(j_N+1)I_{2j_N+1}\right\},
\ee
where $I_{2j+1}$ is the identity matrix of dimension $(2j+1)$.
The labels $j_i$ are uniquely determined from the adjoint representation of the
group $G$. It is an interesting representation-theoretic problem to determine
all labels $j_i$ given a compact simple group $G$. In particular, we will be interested in the
question whether any of the labels $j_i$ can be equal to $1/2$, that is, whether the adjoint
representation of the group $G$ contains the spinor representation of the group $SU(2)$.
We will show below in Sect. 7 that the stability of the model depends on the answer to this
question. We do not solve this question in this paper. We simply carry out the calculations
for arbitrary values of the labels $j_i$.

We will consider spin-tensor fields taking values in the Lie algebra of the gauge group $G$. 
We consider a spin-tensor representation $\Sigma$
of the 
spin group $\Spin(4)$ with generators $\Sigma_{ab}$
satisfying the algebra 
\be
[\Sigma^{ab},\Sigma_{cd}]=-4\delta^{[a}{}_{[c}\Sigma^{b]}{}_{d]}.
\ee
Recall that $\Spin(4)=SU(2)\times SU(2)$. 
That is why the generators
\be
T_i=\frac{1}{2}\varepsilon_{ijk}\Sigma^{jk}
\ee
form a representation $T$ of the subalgebra $su(2)$
\be
[T_i,T_j]=-\varepsilon^{k}{}_{ij}T_k.
\ee
Then the matrices
\be
G_{i}=T_i\otimes I_X+I_T\otimes X_i\,
\ee
form the twisted representation  $T\otimes X$
of the algebra $su(2)$ and satisfy the same algebra
\be
[G_i,G_j]=-\varepsilon^{k}{}_{ij}G_k.
\ee

The covariant derivative acting on spin-tensor fields taking values in the Lie algebra of the group $G$ is defined as follows.
Let $T$ be a general spin-tensor representation of the group $SO(4)$, 
$X$ be a representation of the gauge group $G$ and $\omega^{ab}{}_\mu$ be the spin connection 
corresponding to the chosen orthonormal frame. Then
the covariant derivative is defined by
\be
\nabla^{T\otimes X}_\mu=
\partial_\mu+\frac{1}{2}\omega^{ab}{}_\mu T(\Sigma_{ab})
+X({\cal A}_\mu)
\,.
\ee

\section{Effective Action}
\setcounter{equation}0

There exists a gauge such that
the one-loop effective action of the Yang-Mills theory
is given by
\cite{avramidi95a,avramidi99}
\be
\Gamma_{(1)} = \frac{1}{2}\;\left(\log\Det L_1
-2\log\Det L_0\right)\,,
\ee
where
 $\Det$ is the functional determinant, the operators $L_0$ and $L_1$ have
the form
\be
L_0=-\Delta_{X}\,,
\qquad
L_1=-\Delta_{T_1\otimes X}+Q\,,
\label{33xx}
\ee
$\Delta_{X}=g^{\mu\nu}\nabla^X_\mu\nabla^X_\nu$ is the Laplacian acting on scalar fields in the
representation $X$ of $SU(2)$, 
$\Delta_{T_1\otimes X}=g^{\mu\nu}\nabla^{T_1\times X}_\mu\nabla^{T_1\otimes X}_\nu$ 
is the Laplacian acting on vector fields
in the representation $X$ and $Q$ is an endomorphism defined by
\be
(Q\varphi)^a=\left(R^a{}_b-2{\cal F}^a{}_b\right)\varphi^b
\,,
\ee
where $R^a{}_b$ is the Ricci tensor.

The most important observation that should be made at this point is that the 
{\it positive curvature acts as a mass (or positive potential) term in the Yang-Mills operator}
$L_1$. 
While the magnetic field reduces the eigenvalues of the Yang-Mills operator
 the positive Ricci tensor increases them. Roughly speaking,
it is the balance of these two terms that determines whether or not the Yang-Mills 
operator is positive. However, as we have seen above, the relative value of these two terms
is not arbitrary, {\it one cannot increase the spatial curvature without increasing the chromomagnetic field}
at the same time. It is precisely this feature that leads to the instability of the model, as shown below.

In order to study the infrared behavior of the system, one has to 
introduce an infrared regularization
as in 
\cite{avramidi99}. 
That is why we 
introduce a sufficiently large mass parameter $z$
so that all operators are positive, which is equivalent to replacing the operators $L$ by $L+z^2$.
The determinants of  positive
elliptic operators can be regularized as follows
\cite{avramidi91,avramidi10b}.
We  denote the heat kernel diagonal of the operator $L$ by 
$
U_L(t)
$
and its fiber trace $\Theta_L(t)=\tr U_L(t)$. 
We introduce an arbitrary mass parameter $\lambda$ and define
the coefficients $A^L_k(\lambda)$ by the expansion as $t\to 0$
\be
\Theta_L(t)\sim (4\pi t)^{-2}e^{-t\lambda^2}\sum_{k=0}^\infty
A^L_k(\lambda) t^{k}\,.
\label{322xx}
\ee
Then one can define the renormalized determinant by
\bea
\log\Det_{\rm ren}(L+z^2)
&=&-\vol(M)\int_0^\infty \frac{dt}{t}e^{-tz^2}\;
\Theta^{\rm ren}_L(t)\,,
\eea
where
\bea
\Theta^{\rm ren}_L(t)
&=&\Theta_L(t)
-(4\pi t)^{-2}e^{-t\lambda^2}\left[A^L_0(\lambda)
+A^L_1(\lambda)t
+A^L_2(\lambda)t^{2}\right]\,.
\label{323x}
\eea

Next, let
\bea
\Theta_{YM}(t) &=& \Theta_{L_1}(t)-2\Theta_{L_0},
\\
A^{YM}_k(\lambda)&=&A^{L_1}_k(\lambda)-2A^{L_0}_k(\lambda).
\eea
Then the effective action has the form \cite{avramidi12}
\bea
\Gamma^{}_{(1)}&=&
-\frac{1}{2}\vol(M)\left\{
\beta_{}\log\frac{\mu^2}{\lambda^2}
+
\int_0^\infty \frac{dt}{t}\;e^{-tz^2}
\Theta^{\rm ren}_{YM}(t)
\right\}
\,,
\label{332}
\eea
where $\beta=A^{YM}_2(0)-z^2A^{YM}_1(0)+\frac{1}{2}z^4A^{YM}_0(0)$,
\be
\fl
\Theta^{\rm ren}_{YM}(t)
=
\Theta_{YM}(t)
-(4\pi t)^{-2}e^{-t\lambda^2}\left[A^{YM}_0(\lambda)
+A^{YM}_1(\lambda)t
+A^{YM}_2(\lambda)t^{2}\right].
\label{323xx}
\ee

Finally, there are two simplifications we can make. First, the heat kernel for a Laplace type operator 
$L=-\Delta+Q$ with a covariantly constant potential $Q$ factorizes
\be
\exp(-tL)=\exp(-tQ)\exp(t\Delta)\,.
\ee
Second,  for the product manifold $M=S^1\times S^3$ the heat kernel
for the Laplacian factorizes accordingly
\be
\exp(t\Delta)=\exp(t\partial_\tau^2)\exp(t\Delta_{S^3}),
\ee
that is,
\be
U_L(t)=U^{S^1}(t)\exp(-tQ)
U^{S^3}(t),
\ee 
where $U^{S^1}(t)$ and $U^{S^3}(t)$ are the heat kernel diagonals
for the Laplacians on $S^1$ and $S^3$.
Thus, we obtain finally
\bea
\Theta_{YM}(t)
&=&U^{S^1}(t)
\tr_{X}\Big[\tr_{T_1}\exp(-tQ)
U^{S^3}_{T_1\otimes X}(t)
-2U^{S^3}_{X}(t)\Big]\,,
\eea
where $U^{S^3}_{T_1\otimes X}(t)$ and
$U^{S^3}_{X}(t)$ are the heat kernel diagonals of the Laplacians
acting on vectors and scalars on $S^3$ in the representation $X$.

\section{Heat Kernel on $S^1$}
\setcounter{equation}{0}

We denote the radius of the circle $S^1$ by $a_1$.
There are two dual representations of the heat kernel diagonal: the spectral one
\be
U^{S^1}(t)=\frac{1}{2\pi a_1}
\sum_{n=-\infty}^\infty \exp\left(-\frac{t}{a_1^2}n^2\right),
\ee
and the geometric one
\be
U^{S^1}(t)=(4\pi t)^{-1/2}
\sum_{n=-\infty}^\infty
\exp\left(-\frac{a_1^2 \pi^2}{t}n^2\right).
\ee
That is why we will write it in the form
\be
U^{S^1}(t)=(4\pi t)^{-1/2}\Omega\left(\frac{t}{a_1^2}\right),
\ee
where the function $\Omega(t)$ is defined by
\be
\Omega(t)
=\sum_{n=-\infty}^\infty
\exp\left(-\frac{n^2\pi^2}{t}\right)
=\theta_3\left(0,e^{-\pi^2/t}\right),
\ee
and $\theta_3(v,q)$ is the third Jacobi theta function.
This function satisfies the Poisson duality formula
\be
\Omega(t)
=\sqrt{\frac{t}{\pi}}
\Omega\left(\frac{\pi^2}{t}\right)\,
\ee
and has the following asymptotics: as $t\to 0$
\be
\Omega(t)=1+2e^{-\pi^2/t}+O\left(e^{-4\pi^2/t}\right)
\label{46}
\ee
and as $t\to\infty$
\be
\Omega(t)=\sqrt{\frac{t}{\pi}}\left[1+2e^{-t}+O\left(e^{-4t}\right)\right].
\ee

\section{Geometry of $S^3$}
\setcounter{equation}0

 We consider the sphere $S^3$ of radius $a$.
However, to simplify notation we set the radius $a=1$.
It can be easily reintroduced later from dimensional arguments.
Let $x^i$ be the normal geodesic coordinates on $S^3$ with the origin at the North pole and
ranging
over $(-\pi,\pi)$. In this section all indices denote tensor components
on $S^3$.
The position of the indices on the coordinates will be irrelevant,
that is,
$
x_i=x^i\,.
$
We introduce the radial coordinate
$
{r}=|x|=\sqrt{x_ix^i}\,,
$
and
the angular coordinates (the coordinates on $S^2$)
$
{\theta^i}=\hat x^i=x^i/r\,,
$
(so that $\theta_i\theta^i=1$).
The round metric on $S^3$ in geodesic coordinates is
\be
ds^2=dr^2+\sin^2r\, d\theta^i d\theta_i.
\ee
That is, the metric tensor is
\be
g_{ij}=\theta_i\theta_j+\frac{\sin^2 r}{r^2}\Pi_{ij},
\ee
where
\be
\Pi_{ij}=\delta_{ij}-\theta_i\theta_j\,.
\ee

The sphere $S^3$ can be viewed as the homogeneous space $S^3=SO(4)/SO(3)$
or $S^3=[SU(2)\times SU(2)]/SU(2)=SU(2)$. The geodesic coordinates on $S^3$ are exactly the
canonical coordinates on the group $SU(2)$.
Let $F(q,p)$ be the group multiplication map in canonical coordinates.
This map has a number of important properties.
In particular,
$F(0,p)=F(p,0)=p$ and $F(p,-p)=0$.
Another obvious but very useful property of the group map is that if $q=F(\omega,p)$,
then $\omega=F(q,-p)$ and $p=F(-\omega,q)$. 
Also, there is the associativity
property
$
F(\omega, F(p,q))=F(F(\omega,p),q)\,
$
and the inverse property
$
F(-p,-\omega)=-F(\omega,p).
$

Let $C_i$ be the matrices of the adjoint representation of $su(2)$ defined by
\be
(C_i)^j{}_k=-2\varepsilon^j{}_{ik};
\ee
they satisfy the algebra
\be
[C_i, C_j]=-2\varepsilon^k{}_{ij}C_k.
\ee
Note that the Casimir operator in this normalization is $C_aC_a=-8\, I$.
Let $C=C(x)$ be the matrix defined by
\be
C=C_i x^i\,,
\ee
and let $R$, $Y$ and $D$ be the matrices defined by
\bea
Y=\frac{\II-\exp(-C)}{C}\,,\qquad
R=\frac{C}{\II-\exp(-C)}\,, \qquad
D=\exp C\,.
\eea
Obviously, $RY=YR=I$ and $D=Y^TR$.

To compute these matrices explicitly we note that
\be
C^2=-4r^2\Pi\,,
\ee
and, therefore, 
\be
C^{2n}=(-1)^n(2r)^{2n}\Pi
\,,\qquad
C^{2n+1}=(-1)^n(2r)^{2n} C.
\ee
Therefore, 
the eigenvalues of the matrix $C$ are $(2ir,-2ir,0)$,
and for any analytic function
\be
\fl
f(C)=
f(0)(I-\Pi)
+\Pi \frac{1}{2}\left[f(2ir)+f(-2ir)\right]
+\frac{1}{4ir}C\left[f(2ir)-f(-2ir\right],
\ee
and, therefore,
\bea
\tr f(C)&=&f(0)+f(2ir)+f(-2ir),
\\
\det f(C)&=&f(0)f(2ir)f(-2ir)\,.
\eea
This enables one to compute
\bea
Y&=&I-\Pi+\frac{\sin r}{r}\cos r \Pi
-\frac{1}{2}\frac{\sin^2r}{r^2}C,
\\
R&=&
I-\Pi+r\cot r \Pi+\frac{1}{2}C,
\\
D&=&I-\Pi+\Pi \cos(2r)+\frac{\sin(2r)}{2r}C.
\eea

These matrices satisfy the identities
\be
g_{ij}=\delta_{ab}Y^a{}_i Y^b{}_j, \qquad
g^{ij}=\delta^{ab}R^i{}_a R^j{}_b\,.
\ee
Therefore, both sets of one-forms
\be
\sigma_+^a=Y^a{}_i(x) dx^i, \qquad
\sigma_-^a=Y_i{}^a(x) dx^i,
\ee
form orthonormal bases on the cotangent bundle of $S^3$
and the standard metric on $S^3$ is bi-invariant.
These are the left-invariant and the right-invariant one-forms 
satisfying the identities
\be
d\sigma_\pm^a
=\pm\varepsilon^a{}_{bc}\sigma_\pm^b\wedge \sigma_\pm^c\,.
\ee
To avoid any confusion we will always use the right-invariant orthonormal basis
$\sigma^a_+$.

The right-invariant and the left-invariant vector fields
are defined by
\be
K^-_a=R_a{}^i(x) \frac{\partial}{\partial x^i}\,,\qquad
K^+_a=R^i{}_a(x) \frac{\partial}{\partial x^i}\,.
\ee
They satisfy the algebra
\bea
[K^+_a, K^+_b]&=&-2\varepsilon^c{}_{ab}K^+_c,
\\{}
[K^-_a, K^-_b]&=&2\varepsilon^c{}_{ab}K^-_c,
\\{}
[K^+_a, K^-_b]&=&0,
\eea
and form two mutually commuting representations of the group $SU(2)$.
The Casimir operators of these representations are equal to the scalar Laplacian
\be
\Delta_0 =\delta^{ab}K^-_a K^-_b=\delta^{ab}K^+_a K^+_b\,.
\ee

The left-invariant and the right-invariant vector fields are the Killing vectors
of the metric generating the whole isometry group of the sphere $S^3$.
They provide the orthonormal bases for the tangent bundle. We will always
use the right-invariant orthonormal basis $K^+_a$ dual to $\sigma_+^a$.
Note that the geodesic distance between the points $p$ and $q$ is just equal to $|F(p,q)|$.

The Riemannian volume elements of the metric $g$ is defined as usual
\be
d\vol(x)=g^{1/2}(x)dx^1 \wedge dx^2\wedge dx^3
= \sin^2 r\; dr\;d\theta,
\ee
where $d\theta$ is the volume 
element on $S^2$.
The invariance of the volume element means that
for any fixed $p$
\be
d\vol(x)=d\vol(-x)
=d\vol(F(x,p))=d\vol(F(p,x)).
\ee

We denote the covariant derivative with respect to $K^+_a$
simply by $\nabla_a$.
The Levi-Civita connection of the bi-invariant metric in the right-invariant basis
is defined by
\be
\nabla_{a}K^+_b
=\varepsilon^c{}_{ba}K^+_c,
\ee
so that the coefficients of the affine connection are
\be
\omega^a{}_{bc}
=\sigma_+^a(\nabla_{c}K^+_b)
=\varepsilon^{a}{}_{bc}\,.
\ee
Then
\be
\nabla_{a}K^-_d=
-\varepsilon^b{}_{ca}D_d{}^c
K^+_b.
\ee
Now, the Riemann curvature tensor is
\be
R^a{}_{bcd}=-\varepsilon^f{}_{cd}\varepsilon^{a}{}_{fb}\,,
\ee
and the Ricci curvature tensor 
and the scalar curvature are
\be
R_{ab}=2\delta_{ab}\,,\qquad
R=6\,.
\ee

Let $\nabla$ be  the total connection on a vector bundle $V$
realizing the representation $G$ of the group $SU(2)$. Then 
the Yang-Mills connection on this bundle and its curvature are
\be
{\cal A}=\sigma_+^a G_a, \qquad
{\cal F}=\frac{1}{2}{\cal F}_{ab}\sigma_+^a\wedge \sigma_+^b
=\frac{1}{2}\varepsilon^c{}_{ab}G_c\sigma_+^a\wedge \sigma_+^b.
\ee
The covariant derivative of a section of the bundle $V$ is then
\be
\nabla_a\varphi=(K_a^++G_a)\varphi,
\ee
and the Laplacian takes the form
\be
\Delta=\nabla_a\nabla^a=(K_a^++G_a)(K_a^++G_a)
=\Delta_0+2G^a K_a^++G^2,
\ee
where $G^2=G_iG^i$.
Also, the derivatives along the left-invariant vector fields are
\be
\nabla_{K_a^-}\varphi=(K_a^-+B_a)\varphi,
\ee
where
$
B_a=D_a{}^bG_b.
$

We want to rewrite the Laplacian in terms of Casimir operators of some representations of the group
$SU(2)$. The covariant derivatives $\nabla_a$ do not form a representation of the algebra $SU(2)$. The operators
that do are the covariant Lie derivatives. 
The covariant Lie derivatives along a Killing vector $\xi$ of sections of this vector bundle
are defined by
\be
{\cal L}_\xi=\nabla_\xi-\frac{1}{2}\sigma_+^a(\nabla_{b}\xi)\varepsilon^{bc}{}_a G_c.
\ee
By denoting the Lie derivatives along the Killing vectors $K^\pm_a$ by ${\cal K}^\pm_a$
this gives for the right-invariant and the left-invariant bases
\bea
{\cal K}^+_a&=&{\cal L}_{K^+_a}
=\nabla_{a}+ G_a=K^+_a + 2G_a\,,
\label{536}
\\
{\cal K}^-_a&=&{\cal L}_{K^-_a}
=\nabla_{K^-_a}-B_a=K^-_a\,.
\label{537}
\eea
It is easy to see that these operators form a representation of the isometry
algebra $su(2)\times su(2)$
\bea
[{\cal K}^+_a, {\cal K}^+_b]&=&-2\varepsilon^c{}_{ab}{\cal K}^+_c,
\\{}
[{\cal K}^-_a, {\cal K}^-_b]&=&2\varepsilon^c{}_{ab}{\cal K}^-_c,
\\{}
[{\cal K}^+_a, {\cal K}^-_b]&=&0.
\eea
The Laplacian is given now by the sum of the Casimir operators
\be
\Delta=\frac{1}{2}{\cal K}_+^2+\frac{1}{2}{\cal K}_-^2-G^2,
\label{541x}
\ee
where
${\cal K}_\pm^2={\cal K}^\pm_a{\cal K}^\pm_a$.

We need to compute the action of isometries on $SU(2)$.
Let $T_a$ be the generators of some representation of 
the group $SU(2)$ satisfying the algebra $su(2)$
\be
[T_a, T_b]=-2\varepsilon^c{}_{ab}T_c,
\label{541}
\ee
and $T(x)=T_a x^a$.
First, one can derive a useful commutation formula
\be
\exp[T(x)]T_b\exp[-T(x)]
=D^a{}_b(x)T_a\,.
\ee
Next, one can show that
\bea
K^+_a \exp[T(x)]&=& \exp [T(x)] T_a,
\\
K^-_a \exp[T(x)]&=& T_a \exp [T(x)].
\eea
Therefore,
\be
\Delta_0\exp[T(x)]= \exp [T(x)] T^2= T^2 \exp [T(x)].
\ee
In particular, for the adjoint representation these formulas take the form
\be
D C_b =D^a{}_b C_a D,\qquad
K^+_a D= D C_a, \qquad
K^-_a D= C_a D,
\ee
\be
\Delta_0 D=-8 D.
\ee
This immediately leads to further important equations
\bea
\fl
\exp[K^+(p)]\exp[T(x)]&=&\exp[T(x)]\exp[T(p)]=\exp\left[T(F(x,p))\right],
\\
\fl
\exp[K^-(q)]\exp[T(x)]&=&\exp[T(q)]\exp[T(x)]=\exp\left[T(F(q,x))\right],
\eea
where $K^\pm(p)=p^aK^\pm_a$.
More generally,
\bea
\fl
\exp[K^+(p)+K^-(q)]\exp[T(x)]&=&\exp[T(q)]\exp[T(x)]\exp[T(p)]
\nonumber\\
&=&
\exp\left[T(F(q,F(x,p)))\right].
\eea

Then
for a scalar function $f(\omega)$ the action of the 
right-invariant and left-invariant vector fields is simply
\be
\exp[K^+(p)]f(x)=f(F(x,p)),\quad
\exp[K^-(q)]f(x)=f(F(q,x)),
\ee
more generally,
\be
\exp[K^+(p)+K^-(q)]f(x)=f(F(q,F(x,p))).
\ee

We introduce the average of the group elements over $S^2$ by
\be
\Lambda_T(r)
=\int_{S^2}\frac{d\theta}{4\pi} \exp[rT(\theta)].
\label{554}
\ee
One can show that this is a group invariant. Therefore, it can be only a function of the Casimir operator 
$T^2$. It is determined by the characters of the irreducible representations. For an irreducible representation $j$,
this is equal to 
\be
\Lambda_j(r)=\frac{1}{2j+1}\sum_{|\mu|\le j}\cos(2 \mu r)I_j\,,
\label{555}
\ee
where the sum goes over integer $\mu$ for integer $j$ and over half-integer $\mu$ for half-integer $j$.
Note that the function 
$\Lambda_j(r)$ is periodic with period $\pi$ for integer $j$ and antiperiodic for half-integer $j$, that is,
for all integer $j$:
\be
\Lambda_j(t,r-\pi n)=\Lambda_j(t,r),
\ee
and for all half-integer $j$:
\be
\Lambda_j(t,r-\pi n)=(-1)^n\Lambda_j(t,r).
\ee
We can combine these formulas by writing
\be
\Lambda_j(t,r-\pi n)=(-1)^{2jn}\Lambda_j(t,r).
\ee
For a general reducible representation it is the direct sum of the irreducible ones
\be
\Lambda_T(r)=\diag\{
\Lambda_{j_1}\oplus \cdots \oplus \Lambda_{j_N}
\}.
\ee

Next, we show that
for any representation $T$ of $SU(2)$ we have the identity
\be
\fl
\exp(tT^2)=
(4\pi t)^{-3/2}e^t
\int_{0}^\infty dr\;\int_{S^2}d\theta\;r\sin r \exp\left(-\frac{r^2}{4t}\right)
\exp[rT(\theta)],
\label{560}
\ee
which can also be written as an integral over $\RR^3$
\be
\exp(tT^2)=
(4\pi t)^{-3/2}e^t
\int_{\RR^3} dx\;r\sin r \exp\left(-\frac{r^2}{4t}\right)
\exp[T(x)],
\ee
where, as usual, $r=|x|$.
First, we compute the following Gaussian integral
\bea
&&(4\pi t)^{-3/2}e^t\int_{0}^\infty dr\;r\sin r \exp\left(-\frac{r^2}{4t}\right)
\,4\pi\,\cos(2\mu r)
\nonumber\\
&&=
\left(\frac{1}{2}+\mu\right)e^{-4\mu(\mu+1)t}
+\left(\frac{1}{2}-\mu\right)e^{-4\mu(\mu-1)t}
\eea
Therefore, for any irreducible representation $j$ of $SU(2)$ we have
\bea
&&(4\pi t)^{-3/2}e^t\int_{0}^\infty dr\;r\sin r \exp\left(-\frac{r^2}{4t}\right)
4\pi \Lambda_j(r)
\nonumber\\
&&=
I_j\frac{1}{2j+1}\sum_{|\mu|\le j}
\left\{\left(\frac{1}{2}+\mu\right)e^{-4\mu(\mu+1)t}+\left(\frac{1}{2}+\mu\right)e^{-4\mu(\mu-1)t}\right\}
\nonumber\\
&&=\exp\left[-4j(j+1)t\right]I_j.
\eea
This is nothing but the eq. (\ref{560}) for an irreducible representation. The general case follows
from this trivially.

We note that the above identity (\ref{560}) can also be written
as an integral over the group $SU(2)$
\be
\exp(tT^2)=
\int_{SU(2)} d\vol(x) U_0(t,x)
\exp[T(x)],
\label{564}
\ee
where
\be
U_0(t,x)
= \sum_{n=-\infty}^\infty
(4\pi t)^{-3/2}e^t
\frac{r+2\pi n}{\sin r}\exp\left[-\frac{(r+2\pi n)^2}{4t}\right].
\label{565}
\ee 
One can show that $U_0(t,x)$ is nothing but the scalar heat kernel
on the group $SU(2)$. 
First of all, it satisfies the initial condition
\be
U_0(0,x)=\delta_{S^3}(x).
\ee
Second, it satisfies the heat equation.
This can be shown either by a direct computation or, more elegantly,
as follows.
We have
\bea
0&=&(\partial_t-T^2)\exp(tT^2)
\nonumber\\
&=&
\int_{SU(2)}d\vol(x)
\left(\partial_t-T^2\right) 
U_0(t,x)\exp [T(x)].
\eea
Now, by using the fact that $T^2\exp[T(x)]=\Delta_0\exp[T(x)]$ and by integrating by parts we get
\be 
\int_{SU(2)}d\vol(x)  
\exp [T(x)]
\left(\partial_t-\Delta_0\right) U_0(t,x)=0,
\ee
which gives
\be
\partial_t U_0= \Delta_0 U_0\,.
\ee

\section{Heat Kernel on $S^3$}
\setcounter{equation}0

Our goal is to evaluate the heat kernel diagonal. 
Since it is constant we can evaluate it at any point, say, 
at the origin. 
That is why, 
we will evaluate the heat kernel when one point is fixed at the origin,
which we will denote simply by $U(t;x)$. 
By using eq (\ref{541x}) we obviously have
\be
\exp(t\Delta)=\exp\left(-tG^2\right)\exp\left[\frac{t}{2}\left({\cal K}_+^2+{\cal K}_-^2\right)\right].
\ee
Therefore, the heat kernel is equal to
\be
U(t,x)=\exp\left(-tG^2\right)\Psi
\left(\frac{t}{2},x\right),
\label{62}
\ee
where $\Psi(t,x)=\exp\left[t\left({\cal K}_+^2+{\cal K}_-^2\right)\right]\delta(x)$ 
is the heat kernel of the operator $\left({\cal K}_+^2+{\cal K}_-^2\right)$.

First of all, we note that the scalar heat kernel
has two important properties: the invariance property
\be
 U_0(t,F(p,q))= U_0(t,F(q,p)),
\ee
and the semigroup property
\be
\int_{SU(2)}d\vol(p) U_0(t,F(p,z))  U_0(s,p)= U_0(t+s,z).
\ee

By using the above method, eq. (\ref{564}),
 with $T={\cal K}_+$ and $T=-{\cal K}_-$ we get
\bea
\exp(t{\cal K}^2_++t{\cal K}_-^2)
&=&
\int\limits_{SU(2)\times SU(2)}
 d\vol(q)\;d\vol(p)\;
 U_0(t,q) U_0(t,p)
\nonumber\\
&&\times
\exp[\mathcal{K}^+(q)-{\cal K}^-(p)].
\eea
By using the form of the operators ${\cal K}^\pm_a$, eqs. (\ref{536})-(\ref{537}), we obtain
\bea
\exp(t{\cal K}^2_++t{\cal K}_-^2)
&=&
\int\limits_{SU(2)\times SU(2)} d\vol(q)\;d\vol(p)\;
 U_0(t,q) U_0(t,p)
\nonumber\\
&&\times
\exp[2G(q)]\exp[K^+(q)-K^-(p)]
\eea
Therefore, by acting on the function $\Psi(s,x)$ we get
\bea
\Psi(t+s;x)&=& 
\int\limits_{SU(2)\times SU(2)}d\vol(q)\;d\vol(p)
 U_0(t,q) U_0(t,p)
\nonumber\\
&&
\times
\exp[2G(q)]
\Psi(s;F(-p,F(x,q))).
\eea
We change the variable $p$ by
$
p=F(x,F(q,-z))
$
so that
$
F(-p,F(x,q))=z\,.
$
Then,
\bea
\Psi(t+s;x)&=&
\int\limits_{SU(2)\times SU(2)}d\vol(z)\;d\vol(q)
 U_0(t,q) U_0(t,F(x,F(q,-z)))
\nonumber
\\
&&\times 
\exp[2G(q)]
\Psi(s;z).
\eea
Now, we take the limit $s\to 0$ and use the fact that $\Psi(0,z)=\delta_{S^3}(z)$
to get
\be
\Psi(t;x)=
\int_{SU(2)}d\vol(q)\;
 U_0(t,q)
 U_0(t,F(q,x))
\exp[2G(q)].
\ee
Finally, we compute the diagonal by setting $x=0$
\bea
\Psi(t;0)&=&
\int_{SU(2)}d\vol(q)\;
 U_0(t,q) U_0(t,q)
\exp[2G(q)].
\eea

Now, by using eqs. (\ref{62}), (\ref{565})
we get the heat kernel diagonal 
\be
U^{S^3}(t)=(4\pi t)^{-3/2}e^t\exp\left[-tG^2\right]S(t),
\ee
where
\bea
S(t)&=&\sum_{n,m=-\infty}^\infty
32\pi \int\limits_{0}^\pi dr \;
(r+2\pi n)(r+2\pi m)\Lambda_G(r)
\nonumber\\
&&\times
\exp\left\{-\frac{1}{2t}\left[(r+2\pi n)^2+(r+2\pi m)^2\right]\right\},
\eea
and $\Lambda_G(r)$ is the average of $\exp[2rG(\theta)]$ over the sphere $S^2$ introduced above, eq.
(\ref{554}).
Note that $\Lambda_G(r)$ is an even function of $r$ which is periodic with the period $2\pi$. 
Therefore, the integral can be extended to the interval $[-\pi,\pi]$.
Thus, the integrand is a periodic function with the period $2\pi$. Therefore, by changing the variables by
$r\mapsto r-2\pi m$ and $n\mapsto n+m$
one summation results in the
integration over the whole $\RR$ and we get
\bea
\fl
S(t)&=&
2t^{-3/2}\sum_{n=-\infty}^\infty
\int\limits_{-\infty}^\infty \frac{dr}{\sqrt{\pi}} \;
(r+2\pi n)r\Lambda_G(r)
\exp\left\{-\frac{1}{2t}\left[(r+2\pi n)^2+r^2\right]\right\}.
\eea
Next, we change variable by $r\mapsto r\sqrt{t}-\pi n$ to get
\bea
\fl
S(t)&=&
\sum_{n=-\infty}^\infty
\exp\left(-\frac{\pi^2 n^2}{t}\right)
\int\limits_{-\infty}^\infty \frac{dr}{\sqrt{\pi }} \;e^{-r^2}
\left(2r^2-2\frac{\pi^2 n^2}{t}\right)\Lambda_G\left(r\sqrt{t}-\pi n\right).
\label{614}
\eea

Let us compute the heat kernel diagonal for an irreducible representation $j$ of $SU(2)$.  
By using the  explicit form of the function $\Lambda_j$ we get
\bea
\fl
U^{S^3}_j(t)=\frac{1}{2j+1}\sum_{|\mu|\le j}(4\pi t)^{-3/2}e^{t[j(j+1)+1]} 
\sum_{n=-\infty}^\infty (-1)^{2jn}
\exp\left(-\frac{\pi^2 n^2}{t}\right) 
\nonumber\\
\times\int\limits_{-\infty}^\infty \frac{dr}{\sqrt{\pi }} \;e^{-r^2}
\left(2r^2-2\frac{\pi^2 n^2}{t}\right)\cos\left(2 \mu r\sqrt{t}\right).
\eea
This integral is Gaussian and can be easily computed
\bea
\fl
U^{S^3}_j(t)=\frac{1}{2j+1}\sum_{|\mu|\le j}(4\pi t)^{-3/2}e^{t[j(j+1)+1-\mu^2]} 
\sum_{n=-\infty}^\infty
(-1)^{2jn}
\exp\left(-\frac{\pi^2 n^2}{t}\right) 
\nonumber\\
\times\left\{1-2\mu^2 t-2\frac{\pi^2n^2}{t}\right\}.
\eea
This formula is {\it one of main results of this paper}.
It gives the heat kernel diagonal of the Laplacian on $S^3$ for an arbitrary irreducible
representation of the group $SU(2)$.

At this point it is convenient to introduce a new function
\be
\Omega_j(t)=\sum_{n=-\infty}^\infty (-1)^{2jn}\exp\left[-\frac{\pi^2n^2}{t}\right].
\ee
Then
\bea
\fl
U^{S^3}_j(t)&=&\frac{1}{2j+1}
\sum_{|\mu|\le j}(4\pi t)^{-3/2}e^{t[j(j+1)+1-\mu^2]} 
\left\{(1-2\mu^2 t)\Omega_j(t)-2t\Omega\rq{}_j(t)\right\}.
\eea

Note that for integer $j$
\be
\Omega_j(t)=\Omega(t)=
\sum_{n=-\infty}^\infty \exp\left[-\frac{\pi^2n^2}{t}\right]
=\theta_3\left(0,e^{-\pi^2/t}\right),
\ee
and for half-integer $j$
\be
\Omega_j(t)=\tilde\Omega(t)
=\sum_{n=-\infty}^\infty (-1)^{n}\exp\left[-\frac{\pi^2n^2}{t}\right]
=\theta_4\left(0,e^{-\pi^2/t}\right).
\ee
By using the duality of the theta-functions we can rewrite these functions in the form
\be
\Omega(t)=
\sqrt{\frac{t}{\pi}}\theta_3(0,e^{-t})
=\sqrt{\frac{t}{\pi}}\sum_{n\in \ZZ} e^{-tn^2},
\label{dual1}
\ee
\be
\tilde\Omega(t)=\sqrt{\frac{t}{\pi}}\theta_2(0,e^{-t})
=\sqrt{\frac{t}{\pi}}\sum_{\nu\in \frac{1}{2}+\ZZ} e^{-t\nu^2},
\label{dual2}
\ee
where the first sum goes over all integers and the second sum goes over
all half-integers. Here, of course, $\theta_i(v,q)$ are Jacobi theta functions.

\section{Yang-Mills Heat Trace}
\setcounter{equation}0

Now we can compute the heat trace for the Yang-Mills theory. By using the results for the heat kernel
(and restoring the radius $a$ of the sphere $S^3$)
we get
\bea
\fl
\Theta_{YM}(t)
=(4\pi t)^{-2}\Omega\left(\frac{t}{a_1^2}\right)
e^{t/a^2}
\tr_X\Biggl\{
\tr_{T_1}\exp\left[-\frac{t}{a^2}\left(G^2+a^2Q\right)\right]S_{G}\left(\frac{t}{a^2}\right)
\nonumber\\
-2\exp\left(-\frac{t}{a^2}X^2\right)S_X\left(\frac{t}{a^2}\right)
\Biggr\}.
\eea
For the vector representation $T_1$ the generator $G_a$ has the form
\be
(G_a)^b{}_c=-\varepsilon^b{}_{ac}\otimes I_X+h^b{}_c X_a,
\ee
and the potential term $Q$ has the form
\be
(a^2Q)^b{}_c=2h^a{}_{b}-2\varepsilon^{b}{}_{ca}X_a.
\ee
The Casimir operator is then
\be
(G^2)^b{}_c=-2h^b{}_c+h^b{}_c X^2+2\varepsilon^b{}_{ca}X_a.
\ee
Therefore,
\be
(G^2+a^2Q)^b{}_c=h^b{}_cX^2.
\ee
Thus the above expression simplifies to
\be
\Theta_{YM}(t)
=(4\pi t)^{-2}\Omega\left(\frac{t}{a_1^2}\right)W\left(\frac{t}{a^2}\right)\,,
\ee
where
\bea
W(t)&=& e^t
\tr_X \exp\left(-tX^2\right)\Biggl\{
\tr_{T_1}S_{G}\left(t\right)
-2S_X\left(t\right)
\Biggr\}.
\eea

To compute the function $S_G$ we need to compute the function $\Lambda_G$, that is, the
trace
$
\tr_{T_1}\exp[2G(q)],
$
where
\be
2G(q)=C(q)\otimes I_X+I_{C}\otimes 2X(q).
\ee
Obviously, these two matrices commute. Therefore,
\be
\tr_{T_1}\exp[2G(q)]=\exp[2X(q)]\tr_{T_1}\exp[C(q)].
\ee
The eigenvalues of the matrix $C(q)$ are $(0,0,2i|q|,-2i|q|)$;
therefore,
\be
\tr_{T_1}\exp[C(q)]=2+2\cos(2|q|),
\ee
and we obtain the trace
\be
\tr_{T_1}\exp[2G(q)]=2\left[1+\cos\left(2|q|\right)\right]\exp[2X(q)].
\ee

Now, by using the form (\ref{614}) of the function $S(t)$ we obtain
\bea
W(t)&=&e^t
\sum_{n=-\infty}^\infty 
\exp\left(-\frac{\pi^2 n^2}{t}\right)
\int\limits_{-\infty}^\infty \frac{dr}{\sqrt{\pi }} \;e^{-r^2}
\left(2r^2-2\frac{\pi^2 n^2}{t}\right)
\nonumber\\
&&
\times
2\cos\left(2r\sqrt{t}\right)
\tr_X \exp\left(-tX^2\right)\Lambda_X\left(r\sqrt{t}\right).
\eea
Next, by using the explicit form of the function $\Lambda_X$ in terms of the irreducible representations
(\ref{555})
we get
\bea
W(t)&=&
\sum_{i=1}^N
\sum_{|\mu|\le j_i}
e^{[j_i(j_i+1)+1]t}
\sum_{n=-\infty}^\infty (-1)^{2j_in}
\exp\left(-\frac{\pi^2 n^2}{t}\right)
\\
&&
\times
\int\limits_{-\infty}^\infty \frac{dr}{\sqrt{\pi }} \;e^{-r^2}
\left(2r^2-2\frac{\pi^2 n^2}{t}\right)
2\cos\left(2r\sqrt{t}\right)\cos\left(2\mu r\sqrt{t}\right).
\nonumber
\eea
Now the integrals over $r$ are Gaussian and can be computed;
we obtain finally
\bea
\fl
W(t)=
\sum_{i=1}^N \sum_{|\mu|\le j_i}e^{[j_i(j_i+1)+1]t}
\sum_{n=-\infty}^\infty (-1)^{2j_in}
\exp\left(-\frac{\pi^2 n^2}{t}\right)
\\
\fl
\times
\Bigg[\left(1-2t(\mu+1)^2-2\frac{\pi^2n^2}{t}\right)e^{-t(\mu+1)^2}
+\left(1-2t(\mu-1)^2-2\frac{\pi^2n^2}{t}\right)e^{-t(\mu-1)^2}\Bigg].
\nonumber
\eea
Also, one can express the function $W$ in terms of the function $\Omega_j$ introduced above
\be
W(t)=\sum_{i=1}^N W_{j_i}(t),
\ee
where
\bea
\fl
W_j(t)=
\sum_{|\mu|\le j}e^{[j(j+1)+1]t}
\Bigg\{\left[\left(1-2t(\mu+1)^2\right)
\Omega_{j}(t)-2t\Omega_{j}\rq{}(t)\right]e^{-t(\mu+1)^2}
\nonumber\\
+\left[\left(1-2t(\mu-1)^2\right)\Omega_{j}(t)
-2t\Omega_{j}\rq{}(t)\right]e^{-t(\mu-1)^2}\Bigg\}.
\eea

We will need the asymptotics of the function $W_j(t)$ as $t\to 0$ and as $t\to \infty$.
The asymptotics of the function $W_j(t)$
as $t\to 0$ are
\be
W_j(t)\sim (4j+2)-(8j+4)t+\left(\frac{22}{3}j^3+11j^2+\frac{11}{3}j\right)t^2+\dots\,.
\ee
By using the dual representation (\ref{dual1}), (\ref{dual2}), 
of the functions $\Omega_j(t)$
we can rewrite the function $W_j$ in the form convenient for the calculation of the
asymptotics as $t\to \infty$. After some straightforward but tedious cancellations
the result takes the form
\bea
\fl
W_j(t)=\frac{2}{\sqrt{\pi}}t^{3/2}e^{[j(j+1)+1]t}
\sum_{|\mu|\le j<|\nu|} 
\Bigg\{\left[(\nu+1)^2-(\mu+1)^2\right]e^{-t[(\nu+1)^2+(\mu+1)^2]}
\nonumber\\
+\left[(\nu-1)^2-(\mu-1)^2\right]e^{-t[(\nu-1)^2+(\mu-1)^2]}
\Bigg\},
\eea
where the summation goes over integer $\mu$ and $\nu$ for integer $j$
and over half-integer $\mu$ and $\nu$ for half-integer $j$ (the sum over $\mu$ is finite
and the sum over $\nu$ is infinite).

Thus, as $t\to \infty$ we obtain: for integer $j$
\be
W_j(t)\sim \frac{4}{\sqrt{\pi}} j^2 t^{3/2}e^{-t\lambda^{\rm min}_j},
\ee
where 
\be
\lambda^{\rm min}_j=-j-1,
\ee
and for half-integer $j\ge 3/2$
\be
W_j(t)\sim \frac{4}{\sqrt{\pi}} \left(j^2-\frac{1}{4}\right)
t^{3/2}e^{-t\lambda^{\rm min}_j},
\ee
where
\be
\lambda^{\rm min}_j=-j-\frac{3}{4}.
\ee
Thus, for integer $j\ge 1$
\be
\lambda^{\rm min}_j\le -2\,.
\ee
Also, for any half-integer $j\ge 3/2$,
\be
\lambda^{\rm min}_j\le -\frac{9}{4}\,.
\ee

The only case when the minimal eigenvalue is positive is when $j=1/2$;
we show below that in this case
\be
\lambda^{\rm min}_{1/2}=\frac{19}{4}.
\ee
This is a {\it one of the main results of this paper}. 
It tells us that the {\it minimal eigenvalue is positive only in the case
when the representation $X$ does not contain any higher spin representations
of $SU(2)$ with $j\ge 1$
but contains only the spinor (fundamental) representation} of $SU(2)$ with $j=1/2$.
In all other cases, that is, {\it when the representation $X$ contains at least one representation with $j\ge 1$, 
the minimal eigenvalue is negative and the 
heat kernel grows exponentially at infinity, at least as 
$e^{2t}$, leading to the infrared instability of the chromomagnetic vacuum}. 

Now, we can compute the effective action. First, by using the asymptotics of the function $W_j(t)$
as $t\to 0$ we get
the asymptotics of the heat trace
\be
\Theta_{YM}(t)\sim (4\pi t)^{-2}\left\{C_0+C_1 
\frac{t}{a^2}+C_2\frac{t^2}{a^4}+\cdots \right\},
\ee
where
\bea
C_0&=& \sum_{i=1}^N(4j_i+2),
\\
C_1&=&  \sum_{i=1}^N(-8j_i-4),
\\
C_2&=&  \sum_{i=1}^N\left(\frac{22}{3}j_i^3+11j_i^2+\frac{11}{3}j_i
\right).
\eea
Therefore,
$
\beta=(4\pi )^{-2}\left[C_2-z^2a^2C_1+\frac{1}{2}z^4a^4C_0\right]/a^4.
$
Now, the renormalized heat trace becomes
\be
\Theta_{YM}^{\rm ren}(t)=(4\pi t)^{-2}\left\{\Omega\left(\frac{t}{a_1^2}\right)
W\left(\frac{t}{a^2}\right)-R_{YM}\left(\frac{t}{a^2}\right)\right\},
\ee
where
\bea
\fl
R_{YM}\left(t\right)=e^{-t\lambda^2}\left\{
C_0+\left(C_1+C_0\lambda^2\right) t
+\left(C_2+C_1\lambda^2+\frac{1}{2}C_0\lambda^4\right) t^2\right\}.
\eea

Thus, the one-loop effective action is
\bea
\Gamma^{}_{(1)}&=&
-\frac{\pi}{8x}\left\{
\left[C_2-z^2a^2C_1+\frac{1}{2}z^4a^4C_0\right]\log \frac{\mu^2}{\lambda^2}
+\Phi(x)
\right\}
\,,
\eea
where 
$
x=a/a_1
$
and (after rescaling the integration variable $t\to ta^2$)
\be
\Phi(x)=
\int_0^\infty \frac{dt}{t^3}\;e^{-ta^2z^2}
\left\{\Omega\left(x^2t\right)
W\left(t\right)-R_{YM}\left(t\right)\right\}.
\label{746aa}
\ee
The total effective action (including the classical action) is then
\be
\Gamma=
\frac{\pi}{8x}\left\{
\frac{8\pi^2\sigma}{e^2}
-\left[C_2-z^2a^2C_1+\frac{1}{2}z^4a^4C_0\right] \log\frac{\mu^2}{\lambda^2}
-\Phi(x)
\right\},
\ee
where $\sigma=-\tr X^2=\sum_{i=1}^N j_i(j_i+1)$.
This formula is another {\it important result of this paper}.
It gives an exact integral representation of the infra-red regularized
one-loop effective action for pure Yang-Mills theory in Einstein universe.

Recall that $z$ is an infrared regularization parameter introduced to ensure
convergence of the integral (\ref{746aa}) at infinity.
Eventually we need to take off the infrared regularization, that is,
to take the limit $z\to 0$. The convergence of this integral in this limit
depends on the asymptotic properties of the function $W(t)$ at infinity.
By using the asymptotics of the function $W(t)$ at infinity obtained above
we see that as a function of the infrared regularization parameter $z$
the effective action $\Gamma(z)$ is analytic for ${\rm Re}\, z^2> -\lambda_{\rm min}/a^2$,
where
\be
\lambda_{\rm min}=\min_{1\le i\le N}\lambda^{\rm min}_{j_i}\,,
\ee
and has a branch singularity at $z=\sqrt{-\lambda_{\rm min}}/a$.
Therefore, if $\lambda_{\rm min}>0$ the effective action has a well defined regular value
as $z\to 0$, but if $\lambda_{\rm min}<0$
the effective action is singular in this limit and the model is unstable at low energies.

Recall that the stable configuration only occurs when the adjoint representation of the gauge group
$G$ contains only the spinor representation of $SU(2)$.
In principle, for any given group $G$ containing the group $SU(2)$ as a subgroup
we should be able to determine all the labels $j_i$ of the representation $X$, which might be an interesting
representation-theoretic problem.
If one could show that among the labels $j_i$ there cannot be any label $j=1/2$ then, we would
simply conclude that the model is unstable for any compact simple group, a very
strong assertion. 

However, we do not do this in this paper. We will simply assume that this is possible and proceed 
with the calculation.
That is why we will consider further the case when the representation $X$
contains only spinor representations, that is, all $j_i=1/2$. 
In this case the function $W$ simplifies
\bea
\fl
W(t)=N
\sum_{n=-\infty}^\infty (-1)^{n}
e^{-\pi^2n^2/t}
\Bigg[\left(2-9t-4\frac{\pi^2n^2}{t}\right)e^{-t/2}
+\left(2-t-4\frac{\pi^2n^2}{t}\right)e^{t3/2}\Bigg]
\nonumber\\
\fl
=N\left\{\left[\left(2-9t\right)\tilde\Omega(t)-4t\tilde\Omega\rq{}(t)\right]e^{-t/2}
+\left[\left(2-t\right)\tilde\Omega(t)-4t\tilde\Omega\rq{}(t)\right]e^{t3/2}
\right\}.
\eea
Therefore, as $t\to 0$
\bea
W(t)&\sim&
 N\left(4-8t+\frac{11}{2}t^2+\cdots\right),
\eea
so that
\be
C_0=4N\,,\qquad
C_1=-8N, \qquad
C_2=\frac{11}{2}N.
\ee
Further, by using the dual representation of the function $\tilde\Omega$ we obtain
\bea
\fl
W(t)&=&N
\frac{8}{\sqrt{\pi}}t^{3/2}\sum_{n=0}^\infty e^{-t(n+1/2)^2}
\Biggl\{(n-1)(n+2)e^{-t/2}
+n(n+1)e^{t3/2}
\Biggr\}.
\eea
Therefore, as $t\to \infty$
\bea
W(t)&\sim &\frac{48}{\sqrt{\pi}}Nt^{3/2} e^{-19t/4}.
\eea

It is convenient to introduce the function
\be
\tilde W(t)=\frac{1}{N}e^{t 19/4 }W(t).
\ee
Then as $t\to 0$
\be
\tilde W(t)\sim
4+11 t+\frac{101}{8}t^2+\cdots,
\ee
and as $t\to\infty$
\be
\tilde W(t)\sim
\frac{48}{\sqrt{\pi}}t^{3/2} 
+\cdots.
\label{762}
\ee

In this case the effective action is well
defined 
even in the limit when the infrared regularization parameter $z$ is set to zero.
Therefore, by taking the limit $z\to 0$ and choosing the parameter
$\lambda$ by $\lambda=\sqrt{19}/(2a)$ we obtain
\be
\Gamma=
\frac{\pi}{8x}N\left\{
\frac{6\pi^2}{e^2}
-\frac{11}{2}\log \left(\frac{4}{19}a^2\mu^2\right)
-\tilde\Phi(x)
\right\},
\ee
where
\be
\tilde\Phi(x)=
\int_0^\infty \frac{dt}{t^3}\;e^{-t19/4}
\left\{\Omega\left(x^2t\right)
\tilde W\left(t\right)-4-11t
-\frac{101}{8}t^2
\right\}.
\label{764a}
\ee
Note that as $x\to 0$
this function approaches a well defined constant
\be
\Phi_0=
\int_0^\infty \frac{dt}{t^3}\;e^{-t19/4}
\left\{\tilde W\left(t\right)-4-11t-\frac{101}{8}t^2
\right\}.
\ee
Therefore, we can split the integral in two parts
\be
\tilde\Phi(x)=\Phi_0+\Phi_1(x),
\ee
where (after rescaling $t\mapsto t/x$)
\be
\Phi_1(x)=
x^2\int_0^\infty \frac{dt}{t^3}\;\exp\left(-\frac{19}{4}\frac{t}{x}\right)
\left[\Omega\left(xt\right)-1\right]
\tilde W\left(\frac{t}{x}\right).
\ee

By using the asymptotics of the function $[\Omega(t)-1]$
it is not difficult to see that there is a critical value
\be
x_c=\frac{\sqrt{19}}{2}\pi.
\ee
If $x<x_c$ then the integral $\Phi_1(x)$ is
exponentially small.
Indeed, in the limit $x\to 0$ the function $[\Omega(xt)-1]$ is determined by 
the first exponential term, (\ref{46}); therefore,
\be
\Phi_1(x)\sim 2x^2\int_0^\infty \frac{dt}{t^3}\exp\left[-\frac{1}{x}\left(\frac{19}{4}t+\frac{\pi^2}{t}\right)\right]
\tilde W\left(\frac{t}{x}\right).
\ee
The main contribution to this integral comes from the neighborhood of the point $t_0=2\pi/\sqrt{19}$.
Therefore, by using the asymptotics of the function $\tilde W(t)$ (\ref{762})
we obtain
\be
\Phi_1(x)\sim \frac{96}{\pi} x^{5/2}
\exp\left(-\frac{\sqrt{19}\,\pi}{x}\right).
\ee

By rewriting the integral (\ref{764a}) in the form
\bea
\fl
\tilde\Phi(x)=
x^4\int_0^\infty \frac{dt}{t^3}\;e^{-t19/(4x^2)}
\left\{\Omega\left(t\right)
\tilde W\left(\frac{t}{x^2}\right)-4-11\frac{t}{x^2}
-\frac{101}{8}\frac{t^2}{x^4}
\right\},
\label{771}
\eea
one can obtain the asymptotics of the function $\Phi(x)$ as $x\to \infty$
\be
\tilde\Phi(x)
\sim \nu x^4\,,
\ee
where $\nu$ is a positive constant defined by
\be
\nu=4 \int_0^{\infty} \frac{dt}{t^3}
\left[\Omega\left(t\right)-1\right]\,.
\ee
It can be computed exactly $\nu=\frac{8}{\pi^4}\zeta(4)=\frac{8}{90}$,
with $\zeta(s)$ the Riemann zeta function. Thus,
\be
\tilde\Phi(x)
\sim \frac{8}{90} x^4\,.
\ee
By expanding the function $\tilde W(t)$ in a power series one can also obtain the expansion
of the function $\tilde\Phi(x)$ in inverse powers of $x$.

\section{Thermodynamics of Yang-Mills Theory}
\setcounter{equation}0

In this section we investigate the entropy and the heat capacity of the
model for the stable configuration of the background fields containing only one
representation with $j=1/2$.
The temperature is related to the radius of the circle $S^1$
by
$
T=1/(2\pi a_1)=x/(2\pi a)\,.
$
The volume of the space is the volume of the sphere $S^3$,
$V=2\pi^2 a^3
\,.
$
For a canonical statistical ensemble with
fixed $T$ and $V$
the free energy $F=E-TS$ is 
a function of $T$ and $V$ defined by the total effective action
$
F=T\Gamma\,,
$
with $\Gamma=\Gamma_{(0)}+\Gamma_{(1)}$.
Then the entropy, the heat capacity at constant volume 
and the pressure
are determined by the derivatives
of the free energy
\be
S=-\frac{\partial (T\Gamma)}{\partial T}, \qquad
C_v=-T\frac{\partial^2(T\Gamma)}{\partial T^2}\,,\qquad
P=-\frac{\partial (T\Gamma)}{\partial V}
\,.
\ee
It is easy to see that
neither the classical part $\Gamma_{(0)}$
nor the renormalization logarithmic term in the combination $T\Gamma$ 
depend on the temperature.
Therefore,
the entropy and the heat capacity 
do not depend neither on the classical term nor on the renormalization 
parameter $\mu$.
Thus, the entropy and the heat capacity (per unit volume) are given by the derivatives
of the function $\Phi$,
\bea
s=\frac{S}{V}=\frac{N}{16\pi a^3}
\tilde\Phi\rq{}(x)\,,\qquad
c_v=\frac{C_v}{V}=\frac{N}{16\pi a^3}
x\tilde\Phi\rq{}\rq{}(x)\,.
\eea

By using the asymptotics of the function $\Phi(x)$ we can compute now
the entropy and the heat capacity of the gluon gas: 
at low temperature and small volume, as $aT\to 0$, we have
\bea
s&\sim & 
\frac{6\sqrt{38}}{\sqrt{\pi}}Na^{-5/2}T^{1/2}
\exp\left(-\frac{\sqrt{19}}{2aT}\right),
\\
c_v&\sim&
\frac{114}{\sqrt{2\pi}}\, Na^{-7/2}T^{-1/2}
\exp\left(-\frac{\sqrt{19}}{2aT}\right).
\eea
And at high temperature and large volume, as $aT\to \infty$,
we obtain
\bea
s\sim  
\frac{8}{45}\pi^2 NT^3, 
\qquad
c_v\sim
\frac{8}{15}\pi^2 NT^3. 
\eea

One can also compute the energy density $\varepsilon=E/V=T(\Gamma+S)/V$
to get
\be
\varepsilon
=\frac{N}{32\pi^2 a^4}
\left\{
\frac{6\pi^2}{e^2}
-\frac{11}{2}\log  \left(\frac{4}{19}a^2\mu^2\right)
+x\tilde\Phi\rq{}(x)-\tilde\Phi(x)
\right\}\,,
\ee
and the pressure
\be
P
=\frac{N}{96 \pi^2 a^4}\Biggl\{
\frac{6\pi^2}{e^2}-11-\frac{11}{2}\log  \left(\frac{4}{19}a^2\mu^2\right)
+x\tilde\Phi\rq{}(x)-\tilde\Phi(x)
\Biggr\}.
\ee
Notice that both the pressure and the energy density depend
on the renormalization parameter $\mu$.
Further, this gives the equation of state of gluon gas
\be
P=\frac{1}{3}\varepsilon
-\frac{11N}{96 \pi^2 a^4}
\,.
\ee

In the limit of low temperature and small volume as $aT\to 0$ the energy density
has the form
\be
\varepsilon
\sim
\frac{N}{32\pi^2 a^4}
\left\{
\frac{6\pi^2}{e^2}
-\frac{11}{2}\log  \left(\frac{4}{19}a^2\mu^2\right)
-\Phi_0\right\}
\,.
\label{89}
\ee
In the limit of high temperature and large volume, as $aT\to \infty$,
the energy density is
\be
\varepsilon
\sim
\frac{N}{32\pi^2 a^4}
\left\{
\frac{6\pi^2}{e^2}
-\frac{11}{2}\log  \left(\frac{4}{19}a^2\mu^2\right)\right\}
+\frac{2}{15}\pi^2 NT^4\,.
\label{810}
\ee
Of course, in this limit
the gluon gas behaves like the ``colored\rq{}\rq{} 
photon gas that has $2N$  as many degrees of freedom. Recall that for the photon gas
$P=\frac{1}{3}\varepsilon=\frac{1}{45}\pi^2 T^4$ and $s=\frac{4}{45}\pi^2 T^3$.

Notice that at a fixed temperature $T$ as $a\to 0$ the energy density 
$\varepsilon \to +\infty$
and as $a\to \infty$ it goes to a positive constant. Also, both functions
(\ref{89}) and (\ref{810}) have minimums at some different values of the radius $a$,
\be
a_{1,2}=\frac{1}{\mu}b_{1,2}
\exp\left(\frac{12}{11}\frac{\pi^2}{e^2}\right),
\ee
where
\be
b_{1}=\sqrt{\frac{19}{4}}
\exp\left(-\frac{1}{4}\right),\qquad
b_{2}=\sqrt{\frac{19}{4}}
\exp\left(-\frac{1}{4}-\Phi_0\right).
\ee
Therefore, it is reasonable that the energy density has a minimum at a finite
non-trivial value of the radius 
$a_0(T)=\frac{1}{\mu}b(T)
\exp\left(\frac{12}{11}\frac{\pi^2}{e^2}\right)$ between $a_1$ and $a_2$,  
which can be determined numerically.
Obviously, this is a purely non-perturbative effect; it indicates the existence
of a non-trivial vacuum with the magnetic field and the curvature of the order
$\sim\mu^2\exp\left(-\frac{12}{11}\frac{\pi^2}{e^2}\right)$.

\section{Conclusion}

The primary goal of this paper was to study of the low-energy structure of the
Yang-Mills vacuum. We assumed the chromomagnetic nature of the vacuum with
covariantly constant chromomagnetic fields, which was well known to be unstable
in flat space. We noticed that the potential term of the gluon operator
$L_1=-\Delta_{T_1\otimes X}+Q$ has the form
$Q^a{}_b=R^a{}_b-2{\cal F}^a{}_b$. Therefore, a large Ricci curvature increases
the minimal eigenvalue and a large magnetic field decreases it. Therefore, to
make the gluon operator positive one needs a large Ricci tensor and a small
magnetic field. Moreover, one needs the ability to {\it independently} change
the magnitudes of the curvature and the magnetic field to make this work.

Of course, to be able to carry out the calculations one needs some degree of symmetry.
So, we assumed that the chromomagnetic field is covariantly constant and the spatial curvature
is constant and positive. In four dimensions there are not many choices of spaces of non-negative
constant curvature, only products of spheres and circles; the non-flat ones are
$S^1\times S^1\times S^2$ and $S^1\times S^3$. The case of $S^1\times S^1\times S^2$ was
studied in our previous paper \cite{avramidi12}
and in this paper we studied the case of $S^1\times S^3$.

One of our results is the calculation of the 
heat kernel of the Laplacian acting on arbitrary fields on $S^3$ 
in arbitrary representation of $SU(2)$. By using this result one can compute the one-loop 
effective action of any field-theoretic model. We apply it to the calculation of the effective action of 
Yang-Mills theory on Euclidean Einstein universe, $S^1\times S^3$. 
Our main result is the proof that generically the gluon operator almost always has 
negative eigenvalues. The only case in which
the gluon operator does not have negative eigenvalues occurs when
the fundamental (spinor) representation of the group $SU(2)$ can be embedded in the
adjoint representation of the gauge group (that contains the group $SU(2)$ as a subgroup).
If one can show that this is impossible, then our results prove the instability of the chromomagnetic vacuum 
of Yang-Mills theory in Einstein universe for any compact  simple gauge group. We intend to study this representation-theoretic problem in a future work.

We confirm the conslusion of our previous work \cite{avramidi12} that to
stabilize the chromomagnetic vacuum at lower energies one should consider
non-constant magnetic fields on non-compact spaces. Covariantly constant
magnetic fields on compact symmetric spaces are too rigid, they are completely
determined by the spin connection and are of the same order as the spatial
curvature. This makes it impossible for the gluon operator to be strictly
positive.

\ack

The authors would like to thank an anonymous referee for raising the important question about the
embedding of the 
representations of the group $SU(2)$ in the  adjoint representation of the gauge group.


\end{document}